\documentclass[conference]{IEEEtran}
\IEEEoverridecommandlockouts

\usepackage{cite}
\usepackage{amsmath,amssymb,amsfonts}
\usepackage{algorithmic}
\usepackage{graphicx}
\usepackage{textcomp}
\usepackage{xcolor}
\def\BibTeX{{\rm B\kern-.05em{\sc i\kern-.025em b}\kern-.08em
    T\kern-.1667em\lower.7ex\hbox{E}\kern-.125emX}}
\usepackage{algorithm}
\usepackage{array}
\usepackage{url}
\usepackage{verbatim}
\usepackage{tikz}
\usepackage{pgfplots}
\usepackage{filecontents}
\usepackage{siunitx}
\usepackage{svg}

\usepackage[hidelinks]{hyperref}
\usepackage{orcidlink}

\begin{document}

\title{A Quantitative Evaluation of Approximate Softmax Functions for Deep Neural Networks}

\author{
  \IEEEauthorblockN{
    Anthony~Leiva-Valverde\IEEEauthorrefmark{1}\orcidlink{0009-0006-2602-4336},
    Fabricio~Elizondo-Fernández\IEEEauthorrefmark{1},
    Luis~G.~León-Vega\IEEEauthorrefmark{1}\IEEEauthorrefmark{2}\orcidlink{0000-0002-3263-7853},\\
    Cristina~Meinhardt\IEEEauthorrefmark{3},
    Jorge~Castro-God\'{i}nez\IEEEauthorrefmark{1}\orcidlink{0000-0003-4808-4904}
  }
  \vspace{3pt}
  \IEEEauthorblockA{
      \IEEEauthorrefmark{1}Instituto Tecnol\'ogico de Costa Rica, Cartago, Costa Rica\\
      \IEEEauthorrefmark{2}Università degli Studi di Trieste, Trieste, Italia\\
      \IEEEauthorrefmark{3}Universidade Federal de Santa Catarina, Florianópolis, Brasil\\
  }
  \vspace{4pt}
  \IEEEauthorblockA{
    \begin{tabular}{ccc}
      Corresponding authors: & l.leon@tec.ac.cr & jocastro@tec.ac.cr
    \end{tabular}
  }
}
\maketitle

\begin{abstract}
The softmax function is a widely used activation function in the output layers of neural networks, responsible for converting raw scores into class probabilities while introducing essential non-linearity. Implementing Softmax efficiently poses challenges on low-end FPGAs due to limited hardware resources and the computational complexity of exponential and division operations.
This work evaluates approximate computing techniques for softmax acceleration using Taylor series and interpolation methods using Look-Up Tables (LUTs). These approximations aim to reduce execution time and resource consumption while maintaining acceptable levels of numerical precision.
Our findings show that quadratic interpolation with LUTs yields the lowest numerical error. In contrast, Taylor-based approximations offer significantly better performance in terms of execution time and resource efficiency due to their computational simplicity. When applied to real-world deep learning models such as LeNet-5 and MobileNet v2, the first- and second-order Taylor approximations provided substantial trade-offs between accuracy and resource savings, achieving up to 0.2\% accuracy degradation and 14\% resource reduction compared to exact implementations. These results highlight the effectiveness of approximate Softmax designs on resource-constrained FPGAs and lay the groundwork for their integration into larger models, including large language models (LLMs).
\end{abstract}


\section{Introduction}

The softmax function is a version of the logistic function used when having non-binary classifiers. It is often placed at the end of the classifiers as an activation function to extract the probabilities of each output class in a neural network, in particular, after a fully-connected layer (FCL)~\cite{Skansi2018}. A typical example of this usage is in a LeNet-5 model on the MNIST dataset~\cite{lenet}. Apart from its role as a probability extractor, it introduces a non-linearity to the model, enabling the classifications of points with non-linear mappings of data and making the embedded points linearly separable~\cite{nns}, which still applies to state-of-the-art models like Llama 2~\cite{llm-ecaslab}.

In Deep Learning (DL) inference, using 32-bit floating-point (\texttt{float32}) representations provides more precision to the network than required, leading to the concept of \emph{quantisation}: the approximation of the model in other numerical representations with fewer bits~\cite{Liang2021}. Quantisation allows model compression, reducing the memory footprint and better exploitation of vector execution units of CPUs than \texttt{float32}. In the particular case of the activation functions, quantisation mainly accelerates the computation time.

When considering FPGA-based implementations for DL, the vast majority of the implementations for Deep Neural Networks (DNNs) inference are solutions provided by FPGA vendors and open-source initiatives, particularly tailored for high-end FPGAs, such as Xilinx Alveo, Kintex, and Virtex. In those cases where solutions are closed, i.e., no code is available, optimisation possibilities are restricted~\cite{Liang2021}.
%
However, this opens the opportunity to explore solutions based on low-end FPGAs for edge computing, from exploring the synthesis of algorithms to Hardware Description Languages (HDL). High-Level Synthesis (HLS) allows for the implementation of FPGA designs faster than traditional register-transfer level (RTL) descriptions. Moreover, approximate computing techniques can be used for function calculation, possibly having smaller designs with lower power consumption in exchange for numerical accuracy~\cite{Zervakis2015,saadat}.


In this work, we contribute to assessing approximate computing techniques to implement the softmax function using Taylor series and interpolation methods with Look-Up Tables. Each implementation uses Root Mean Square Error (RMSE) to assess numerical error, resource consumption, and impact on actual DL models.

\section{Optimisation framework}

This section presents the function's definition and possible approximations, including Taylor approximation and piece-wise interpolation based on Look-up Tables (LUTs).

\subsection{Definition}

The softmax function is defined as:

\begin{equation}
  \Phi(\mathbf{v})_i = \frac{e^{v_i}}{\sum_{j=1}^k e^{v_j}}
\end{equation}

\noindent
where $v_i$ is the $i$-th element of the input vector $\mathbf{v}$ and $k$ is the number of elements of the vector~\cite{nns}. It involves the computation of the exponential function in a certain domain $S \subset \mathbb{R}$. The domain $S$ can be determined according to the input and output domains of the FCL preceding the softmax function. A FCL is described as the matrix-vector product:

\begin{equation}
    \mathbf{y} = \mathbf{W} \mathbf{x} + \mathbf{b}
\end{equation}

\noindent
where $\mathbf{x}, \mathbf{b}, \mathbf{y}$ are input, bias and output vectors, respectively; and $\mathbf{W}$ is the weights matrix for all the perceptrons within the FCL. For our use case, let us assume a numerical representation that supports an uniformly distributed discrete set within the domain $S = ]-1,1[$, quantised in a fixed-point representation of $\beta$ bits. Hence, an element of the output vector can be expressed as:

\begin{equation}
    y_i = \mathbf{w}_i \cdot \mathbf{x} + b_i
\end{equation}

\noindent
where $\mathbf{w}_i$ is the $i$-th row vector from the matrix $\mathbf{W}$ and $\cdot$ is the dot-product between vectors, expressed as $\mathbf{w}_i \cdot \mathbf{x} = \sum_{j}^{k} w_{ij}x_j$. Each output element involves $k$ products and $k$ additions including the bias. The computation is numerically vulnerable to additions, risking overflows. We can deal with this phenomenon by scaling the operands of the matrix-vector multiplication inversely proportional to the $n$ number of elements of the input vector~\cite{dcas}. Therefore,

\begin{equation}
    y_i = \mathbf{w}_i \cdot \left(\frac{\mathbf{x}}{n}\right) + \frac{b_i}{n}, x_i, w_{ij} \in S \implies y_i \in S
\end{equation}

\noindent
implies that scaling by the inverse of the number of inputs will numerically stabilise the outputs. This is valid under the assumption that the probability distribution just scales numerically without major changes in the shape of the function.

Knowing that the domain of $v_i$ is constrained and given by $S$, the exponential function domain can also be given by $S$. As $S$ is a uniformly distributed discrete set, the function can also be defined by the number of points of the set without incurring an under- or over-discretisation.


\subsection{Taylor approximation}

A Taylor series consists of a function approximation given by the infinite sum of elements expressed in terms of the target function's derivatives at a single point. For the exponential function, the Taylor series centred in $a = 0$ is

\begin{equation}
    e^x = \sum_{n=0}^\infty \frac{x^n}{n!} = 1 + x + \frac{x^2}{2!} + \dots, \forall x \in \mathbb{R}
\end{equation}

\noindent
where $a$ is the point where the function's derivative is centred and it converges everywhere~\cite{taylor}.


\subsection{LUT-based piece-wise interpolation}

Our version of this method consists of sampling the function at uniform, equidistant points and computing the best-fit polynomial between the points. For instance, a linear polynomial requires two points to compute, whereas a quadratic requires three points ~\cite{piecewise}. Fig.~\ref{fig:piecewise} shows how a linear interpolation fits the $e^x$ function by taking eight samples and performing linear interpolation.

\begin{figure}[!t]
    \centering
    \scalebox{0.82}{\begin{tikzpicture}

        \begin{axis}[
            xtick distance=0.5,
            ytick distance=0.5,
            axis equal image=true,
            xlabel={$x$},
            ylabel={$e^x$},
            axis lines=middle,
            xmin=-1, xmax=1,
            ymin=0, ymax=3,
            axis x line=center,
            axis y line=center,
            height=9cm,
		    width=9cm
        ]
            \addplot[color=red, mark=square] table [x=x, y=ex, col sep=comma] {fig/exp.data.csv};
            \addlegendentry{\(\hat{e^x}\)}
            \addplot[color=blue, dashed] {exp(x)};
            \addlegendentry{\(e^x\)}
        \end{axis}
\end{tikzpicture}}
    \caption{Piecewise representation by doing eight samples within the domain $S$ and applying a linear interpolation}
    \label{fig:piecewise}
\end{figure}

The piecewise function segments can be calculated at computation time (runtime) or recalculated at compile time. At runtime, the slope and intercept are computed as 

\begin{equation}
    m_p = \frac{y_{p_1} - y_{p_0}}{x_{p_1} - x_{p_0}}, b_p = y_{p_1} - m_p x_{p_1}
\end{equation}

\noindent such that $f_p(x) = m_p x + b_p, x_{p_0} \le x \le x_{p_1}$, where $(x_{p_0}, y_{p_0}), (x_{p_1}, y_{p_1})$ are the points before and after the point of interest $x_p$, respectively. In this case, the computation of the point requires: (1) storing the points in a LUT, (2) computing the linear equations, and (3) computing the value of interest. Our proposal consists of storing the slope and the intercepts at synthesis time to speed up the computation, shortening the path from (1) to (3).

Moreover, to avoid unwanted divisions while computing the indices of the slope-intercept pairs required for the computation, the number of points can be a power of two, such that the division becomes a bit-shift, in such a way that

\begin{equation}
    p = x^\prime \gg P \implies m_p = M[p], b_p = B[p]
\end{equation}

\noindent where $P$ is the number of points (power of two), $x^\prime$ is the quantised value of $x$ in fixed-point, $M$ and $B$ are the LUTs for the slope and intercept, respectively.

\subsection{Numerical error metric}

To assess the accuracy of the approximate models, we use the Root Mean Square Error (RMSE) metric, which is widely adopted to measure the estimation error \cite{4107991}. Because RMSE is listed as an absolute error metric, it establishes a difference between the exact values and the approximate values, defined as:

\begin{equation}
    \text{RMSE}(\hat{v}) = \left(\frac{1}{N}\sum_{i=1}^{N}(v_i-\hat{v}_i)^2 \right)^\frac {1} {2}
\end{equation}

\noindent where $\hat{v}$ is the approximate vector of the model, $N$ is the vector size, and $v_i$ represents the exact values.  This metric uses the same formula to measure how far the model's predictions are from actual values. Therefore, there is a direct relationship between the accuracy of the model and the value of RMSE.

\begin{figure*}[!t]
    \centering
    \includegraphics[width=0.95\linewidth]{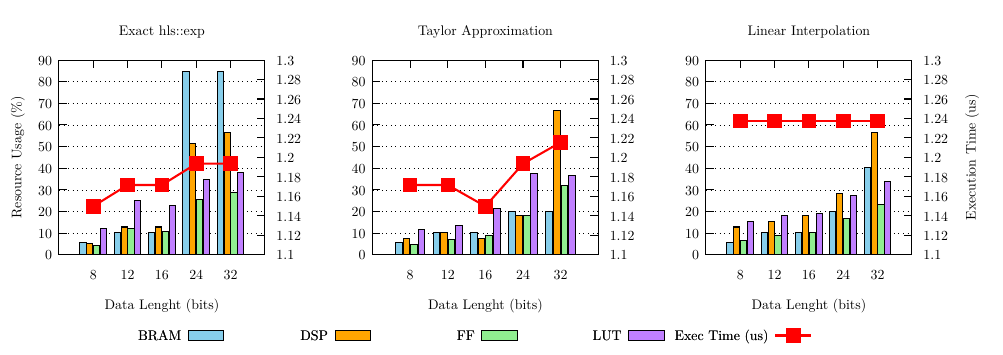}
    \caption{Resource usage and execution time of the softmax accelerators based on 3rd-order Taylor and 64-sample Linear Interpolation processing a 1024 16-bit fixed-point vector. The resource consumption is relative to an AMD Kria KV260.}
    \label{fig:resources-softmax}
    \vspace{-4mm}
\end{figure*}

\section{Standalone Numerical and Performance Evaluation}

In this section, we evaluate different softmax accelerator\footnote{Accelerator's repo: \url{https://github.com/ECASLab/hls-fpga-accelerators/}} configurations implemented on Vitis HLS 2024.01 to observe the difference between them and the exact version provided by Vitis HLS (\texttt{hls::exp}).

Tables \ref{tab:taylor-error} and \ref{tab:interpolation-error} show the error metrics gathered for each softmax approximation type. The results were captured using a test vector with 1000 random values within the softmax domain $S = ]-1,1[$ in a 16-bit fixed-point representation. From all the solutions presented, the approach that generated the lowest error value was quadratic interpolation using LUTs with 64 samples, reaching $\text{RMSE} = \num{2.31e-7}$. In the case of the Taylor approach, the third-order approximation was the one that obtained the best error result with $\text{RMSE} = \num{4.18e-5}$.

\begin{table}[!t]
  \begin{center}
    \caption{Error metrics for the Taylor-Softmax approximation}
    \label{tab:taylor-error}
   \begin{tabular}{| c | c | c | c |}
      \hline
      \textbf{Type} &  \begin{tabular}[c]{@{}l@{}} \textbf{Error}\\ 
      (RMSE) 
      \end{tabular} & \textbf{Variance} & \begin{tabular}[c]{@{}l@{}} \textbf{Standard Deviation} \end{tabular} \\
      \hline
       Order 1 & \num{3.13e-3} & \num{2.48e-6} & \num{1.57e-3} \\
      \hline
       Order 2 & \num{2.97e-3} & \num{2.45e-6} & \num{1.56e-3} \\
      \hline
       Order 3 & \num{4.18e-5} & \num{6.84e-10} & \num{2.62e-5} 
       \\\hline
    \end{tabular}
  \end{center}
  \vspace{-4mm}
\end{table}

\begin{table}[!t]
  \begin{center}
    \caption{Error metrics for the LUT interpolation Softmax with 64 samples}
    \label{tab:interpolation-error}
   \begin{tabular}{| c | c | c | c |}
      \hline
      \textbf{Type} &  \begin{tabular}[c]{@{}l@{}} \textbf{Error} \\ 
      (RMSE) 
      \end{tabular} & \textbf{Variance} & \begin{tabular}[c]{@{}l@{}} \textbf{Standard Deviation} \end{tabular} \\
      \hline
       Lineal & \num{3.22e-6} & \num{4.28e-12} & \num{2.07e-6} \\
      \hline
       Quadratic & \num{2.31e-7} & \num{2.60e-14} & \num{1.61e-7} \\
      \hline
    \end{tabular}
  \end{center}
  \vspace{-4mm}
\end{table}

Regarding the evolution of resource consumption as complexity increases, Fig.~\ref{fig:resources-softmax} shows how the resource consumption and latency scale as the data width changes in softmax accelerators based on both approximation methods. The \emph{Taylor Approximation} uses a third-order Taylor approximation, and the \emph{Linear Interpolation} uses a 64-sample LUT for the exponential function. Both cases use a 16-bit fixed-point data type and a 1024-element vector.

The \emph{Taylor Approximation} shows a latency oscillating between the $1.14~ \mu\text{s}$ and $1.22~\mu\text{s}$, resulting in a faster execution time compared to the \emph{Linear Interpolation}, with a constant execution time of nearly $1.24~\mu\text{s}$ along the different data lengths. The exact version, in contrast, has configurations that are faster than the Taylor approximation, particularly in 8 and 32 bits. In resources, \emph{Taylor Approximation} consumes fewer resources than \emph{Linear Interpolation} up to 16 bits, where the former starts to have less overall consumption than \emph{Taylor Approximation}. In the case of the exact version, the overall consumption is always the greatest. Nevertheless, in both approximate accelerators, the consumption of DSP cells starts to grow exponentially as the data length increases due to the arithmetic complexity involved in the computations.

This highlights a trade-off between the data length, resource consumption, and numerical error. In scenarios where the error resilience is high, the Taylor-based approximation offers a lightweight solution to the computations. Otherwise, the interpolation-based approximation provides a more robust low-error solution that is effective for high-resolution data (16-24-bit fixed-point). Likewise, there are scenarios like the 8-bit configuration, where using the exact version has more benefits than the approximate versions.

\section{Resource and Performance Evaluation on Actual Deep Learning Models}

The key idea behind using approximations in the softmax is to constrain the function domain to reduce the resource consumption in an FPGA. To evaluate the effects on actual DL models, we consider different configurations of the softmax approximated functions at the error level and execution time, in two models: LeNet 5~\cite{lenet}  (evaluated with 10000 samples) and MobileNet v2~\cite{mobilenetv2} (evaluated with 250 samples), using the AxC Executer~\cite{axcexecuter}.

Tables~\ref{tbl:lenet} and~\ref{tbl:mobilenetv2} show the results after accelerating the softmax layer, corresponding to the last layer in both models. In the case of LeNet 5 (Table~\ref{tbl:lenet}), the vector size is 10 elements of 12-bit fixed-point with 6-bit integer part. It shows that the best configuration is the first-order Taylor, with 0.2\% of Top-1 accuracy degradation, keeping the resources low compared to the exact implementation, saving 14\% of resources (with the least saving in FF) with respect to the exact version. On the other hand, for the MobileNet v2 the softmax accelerator processes a 1000-element vector of 20-bit fixed-point elements, with a 10-bit integer part. The Taylor approximation improves the Top-1 accuracy compared to the exact version. This happens because the approximation introduces healthy numerical disturbances within the model, which are not generalisable, as shown in the LeNet-5. second-order Taylor is the best configuration, improving the Top-1 accuracy by 16.6\%, while saving 20\% of overall resources (with the least saving in DSP) with respect to the exact version. The linear interpolation was used to evaluate the exponential function, resulting in a more expensive solution than the Taylor approximation by $1.2\times$ (comparing second-order Taylor and Interpolation of 16 samples).

After evaluating actual DL models with the approximation, it is possible to observe the benefits of approximating the exponential function in the softmax layers. The error resilience of this type of layer is robust enough to support the Taylor approximation, resulting in an opportunity to reduce the computation complexity to a polynomial-like computation. Evaluating more cases such as LLMs may yield interesting results given the amount of softmax computations (the eighth most intensive computation in Llama 2~\cite{llm-ecaslab}).

\begin{table}[!t]
\centering
\caption{LeNet 5 Synthesis Results with 12-bit Fixed-Point (6-bit integer part) and a shift of 3 bits for an AMD Kria KV260}
\begin{tabular}{|l|c|l|l|l|l|}
\hline
\textbf{Configuration}                                             & \multicolumn{1}{l|}{\textbf{\begin{tabular}[c]{@{}l@{}}Top-1\\ Accuracy\end{tabular}}} & \textbf{\begin{tabular}[c]{@{}l@{}}Layer \\ Time (us)\end{tabular}} & \textbf{\begin{tabular}[c]{@{}l@{}}LUT \\ Cells\end{tabular}} & \textbf{\begin{tabular}[c]{@{}l@{}}FF \\ Cells\end{tabular}} & \textbf{\begin{tabular}[c]{@{}l@{}}DSP \\ Cells\end{tabular}} \\ \hline
Exact                                                              & 0.9768                                                                                  & \multicolumn{1}{c|}{0.87}                                                  & \multicolumn{1}{c|}{7940}                                         & \multicolumn{1}{c|}{5362}                                        & \multicolumn{1}{c|}{52}                                         \\ \hline
\begin{tabular}[c]{@{}l@{}}Interpolation\\ 32 samples\end{tabular} & \multicolumn{1}{c|}{0.9763}                                                                                   & \multicolumn{1}{c|}{0.88}                                                                    & \multicolumn{1}{c|}{8050}                                                          & \multicolumn{1}{c|}{5825}                                                         & \multicolumn{1}{c|}{47}                                                            \\ \hline
\begin{tabular}[c]{@{}l@{}}Interpolation\\ 16 samples\end{tabular} & \multicolumn{1}{c|}{0.9763}                                                                                  & \multicolumn{1}{c|}{0.88}                                                                    & \multicolumn{1}{c|}{8010}                                                          & \multicolumn{1}{c|}{5820}                                                         & \multicolumn{1}{c|}{47}                                                            \\ \hline
\begin{tabular}[c]{@{}l@{}}Interpolation\\ 8 samples\end{tabular} & \multicolumn{1}{c|}{0.9765}                                                                                  & \multicolumn{1}{c|}{0.88}                                                                    & \multicolumn{1}{c|}{7997}                                                          & \multicolumn{1}{c|}{5415}                                                         & \multicolumn{1}{c|}{47}                                                            \\ \hline
3rd-order Taylor                                                     & \multicolumn{1}{c|}{0.9763}                                                                         & \multicolumn{1}{c|}{0.89}                                                                    & \multicolumn{1}{c|}{7308}                                                          & \multicolumn{1}{c|}{5879}                                                         & \multicolumn{1}{c|}{52}                                                            \\ \hline
2nd-order Taylor                                                     & \multicolumn{1}{c|}{0.9752}                                                                         & \multicolumn{1}{c|}{0.87}                                                           & \multicolumn{1}{c|}{6684}                                                 & \multicolumn{1}{c|}{4739}                                                & \multicolumn{1}{c|}{42}                                                   \\ \hline
1st-order Taylor                                                     & \multicolumn{1}{c|}{\textbf{0.9751}}                                                                                    & \multicolumn{1}{c|}{\textbf{0.84}}                                                                    & \multicolumn{1}{c|}{\textbf{6544}}                                                          & \multicolumn{1}{c|}{\textbf{4615}}                                                         & \multicolumn{1}{c|}{\textbf{37}}                                                            \\ \hline
\end{tabular}
\label{tbl:lenet}
\end{table}

\begin{table}[!t]
\centering
\caption{MobileNet v2 Synthesis Results with 20-bit Fixed-Point (10-bit integer part) and a shift of 1 bit for an AMD Kria KV260}
\begin{tabular}{|l|c|c|c|c|c|}
\hline
\textbf{Configuration}                                             & \multicolumn{1}{l|}{\textbf{\begin{tabular}[c]{@{}l@{}}Top-1\\ Accuracy\end{tabular}}} & \multicolumn{1}{l|}{\textbf{\begin{tabular}[c]{@{}l@{}}Layer \\ Time (us)\end{tabular}}} & \multicolumn{1}{l|}{\textbf{\begin{tabular}[c]{@{}l@{}}LUT \\ Cells\end{tabular}}} & \multicolumn{1}{l|}{\textbf{\begin{tabular}[c]{@{}l@{}}FF \\ Cells\end{tabular}}} & \multicolumn{1}{l|}{\textbf{\begin{tabular}[c]{@{}l@{}}DSP \\ Cells\end{tabular}}} \\ \hline
Exact                                                              & 0.748                                                                                 & 1.17                                                                                         & 32203                                                                                   & 33043                                                                                  & 160                                                                                   \\ \hline
\begin{tabular}[c]{@{}l@{}}Interpolation\\ 64 samples\end{tabular} & 0.74                                                                                 & 1.19                                                                                   & 28975                                                                              & 26912                                                                             & 128                                                                                \\ \hline
\begin{tabular}[c]{@{}l@{}}Interpolation\\ 32 samples\end{tabular} & 0.688                                                                                 & 1.19                                                                                   & 28847                                                                              & 26816                                                                             & 128                                                                                \\ \hline
\begin{tabular}[c]{@{}l@{}}Interpolation\\ 16 samples\end{tabular}  & 0.556                                                                                 & 1.19                                                                                   & 28559                                                                              & 26752                                                                             & 128                                                                                \\ \hline
3rd-order Taylor                                                     & \textbf{0.872}                                                                                 & 1.17                                                                                   & 37223                                                                              & 37904                                                                             & 224                                                                                \\ \hline
2nd-order Taylor                                                     & \textbf{0.872}                                                                                 & \textbf{1.15}                                                                                   & \textbf{21575}                                                                              & \textbf{22653}                                                                             & \textbf{128}                                                                                \\ \hline
1st-order Taylor                                                     & 0.0                                                                                 & 1.12                                                                          & 18183                                                                     & 20400                                                                    & 64                                                                        \\ \hline
\end{tabular}
\label{tbl:mobilenetv2}
\vspace{-4mm}
\end{table}

\section{Related Work}

Softmax implementations on FPGAs have been explored through both exact and approximate approaches. A fundamental strategy for reducing hardware complexity involves lowering numerical precision by using fixed-point arithmetic representations~\cite{cordic}. Among the exact designs, some works leverage the CORDIC algorithm to compute exponentials and divisions~\cite{cordic, 9531761} efficiently. On the other hand, approximate accelerators aim to reduce computational effort by simplifying multiplication and division operations~\cite{approx-chen}, achieving, for instance, a 3\% accuracy degradation on LeNet-5 using 16-bit fixed-point arithmetic, while consuming only 1354 FFs, 1604 LUTs, and 3 DSP slices, with a latency of $3.788~\text{ns}$. A similar approach is found in~\cite{approx-yao}, which uses a Taylor Series-based approximation, consuming 2229 LUTs and resulting in a 2\% accuracy drop.

In contrast, our design operates with arbitrary precision, illustrating for LeNet-5, 12-bit fixed-point precision. It achieves a significantly lower accuracy degradation—no more than 0.2\%—with a slightly improved delay of $3.65~\text{ns}$, albeit at the cost of approximately four times the area. However, our solution offers a key advantage: it is highly configurable in data precision, order, and number of samples, and it can be tailored to different models and deployment scenarios, providing greater flexibility during development compared to previous works, easily integrable within popular frameworks like hls4ml~\cite{hls4ml}.

\section{Conclusion}
In this work, we explored various approximate implementations of the Softmax function on FPGAs, focusing on Taylor series and linear interpolation with Look-Up Tables (LUTs). Our results indicate that quadratic interpolation offers the lowest numerical error within the softmax domain. However, this method—and linear interpolation more broadly—incurs higher execution times due to the overhead introduced by LUT access and interpolation steps. In contrast, Taylor-based approximations deliver better performance, attributed to their simpler arithmetic structure for approximating the exponential function.

When deployed in real-world deep learning models such as LeNet-5 and MobileNet v2, Taylor approximations proved to be a practical trade-off, introducing minimal accuracy degradation while significantly reducing resource usage on FPGAs. For future work, these approximation techniques show promising potential for accelerating inference in large language models (LLMs), which are increasingly dominant in state-of-the-art AI applications and heavily rely on softmax computations.

\section*{Acknowledgements}
This work was supported by RidgeRun, LLC, and the Costa Rica Institute of Technology under research project 1360058 (Generación Automática de Hardware para Aplicaciones de Aprendizaje Automático basadas en FPGA). Results achieved with the funding obtained under Axis IV of the PON Research and Innovation 2014-2020 "Education and research for recovery - REACT-EU".

\bibliographystyle{IEEEtran}
\bibliography{output}

\end{document}